\documentclass[conference]{IEEEtran}
\IEEEoverridecommandlockouts
\usepackage{cite}
\usepackage{amsmath,amssymb,amsfonts}
\usepackage{algorithmic}
\usepackage{graphicx}
\usepackage{textcomp}
\usepackage{xcolor}
\usepackage{subcaption}
\begin{document}

\title{GeoGS-CE: Learning Delay--Beam Channel Priors with 3D Gaussians for High-Mobility Scenarios
\thanks{This work was supported by the Hong Kong Research Grants Council under the Areas of Excellence scheme grant AoE/E-601/22-R and NSFC/RGC Collaborative Research Scheme grant CRS-HKUST603/22.}
}
\author{\IEEEauthorblockA{Yumeng Zhang, Jiajia Guo, Chaozheng Wen, Chenghong Bian, and Jun Zhang}\IEEEauthorblockA{\textit{iComAI Lab, HKUST}}
\IEEEauthorblockA{\textit{Email:} eeyzhang@ust.hk,eejiajiaguo@ust.hk, cwenae@connect.ust.hk, eechbian@ust.hk, eejzhang@ust.hk}}
\maketitle

\begin{abstract}
Wideband channel estimation (CE) in high-mobility scenarios remains challenging because channel responses vary rapidly, while practical systems can allocate only sparse pilots to accommodate dense users. Fortunately, many high-mobility environments, such as high-speed railways, exhibit scheduled trajectories, predictable velocities, and a limited number of dominant propagation paths. These properties induce a delay--beam power spectrum that is more stable than the instantaneous complex channel frequency response (CFR), less sensitive to the random phase coherence, and rich in geometric information. 
To exploit such environmental properties, we propose \textbf{GeoGS-CE}, a two-stage channel estimation framework for sparse-pilot high-mobility scenarios. In the offline stage, GeoGS-CE jointly models: 1) a scene-level 3D Gaussian representation that captures the non-line-of-sight (NLoS) geometric scattering support, and 2) a leakage-aware differentiable wireless rendering process that maps the NLoS Gaussians, together with an explicit virtual line-of-sight (LoS) component, to the measured delay--beam power spectrum, while accounting for practical OFDM delay and array leakage effects.
 In the online stage, the delay--beam power spectrum is predicted for each user location and used as a strong covariance prior, enabling accurate full-band and full-array CFR reconstruction and tracking through a linear MMSE estimator. Simulations based on channels generated from a segment of the Guangshen high-speed railway show that the proposed geometric prior substantially improves CFR reconstruction over pilot-only and non-geometric baselines.
\end{abstract}


\section{Introduction}
Accurate channel estimation and prediction are essential for wideband massive-MIMO systems, yet they remain challenging in high-mobility scenarios where channel responses vary rapidly. At the same time, practical systems can allocate only limited pilot overhead to each user, which further degrades channel reconstruction performance \cite{zhou2015channel}. Fortunately, many high-mobility scenarios, such as high-speed railways and unmanned aerial vehicles (UAVs), exhibit scheduled trajectories, predictable velocities, and a small number of dominant geometry-dependent propagation paths \cite{wu2021channel,khawaja2016uwb}. These properties disclose the possibility of learning strong channel priors to assist sparse-pilot channel estimation and prediction.

A common solution is to exploit sparsity in transformed delay--beam (or delay-angle) domains \cite{wan2024two}. Model-based sparse recovery methods, including compressed sensing and Bayesian estimators, improve reconstruction under sparse pilots by imposing structured priors \cite{11373228}. However, they often require iterative estimators, careful hyperparameter tuning, and strong initialization, which limits their suitability for real-time reconstruction of rapidly varying channels \cite{11462455}. Learning-based methods help to reduce online complexity, but many directly learn the coherent channel information such as channel frequency response (CFR), making them less interpretable and less robust to changes in pilot patterns, resource allocation, or trajectories \cite{zhao2023nerf2}.

Recently, 3D Gaussian Splatting (3D-GS) has emerged as an efficient and interpretable representation for scene modeling, and has been extended to wireless rendering tasks such as radio-map reconstruction \cite{11258087}, cross-frequency prediction \cite{li2025wideband}, pathloss/angle-spectrum prediction \cite{zhang2026rf}, and efficient wireless rendering pipelines \cite{11044513}. Nevertheless, applying 3D-GS to wideband channel reconstruction remains non-trivial. Direct CFR supervision is highly sensitive to small path-length errors and random phase rotations, making coherent channel prediction difficult, especially in high-mobility wideband systems \cite{khawaja2016uwb}.
Fortunately, in fixed railway corridors, although coherent channel coefficients fluctuate rapidly, the dominant non-coherent delay--beam support is largely governed by the route geometry, surrounding scatterers, and BS array configuration, and therefore evolves more smoothly. This motivates learning a geometry-conditioned non-coherent channel prior rather than directly predicting the full complex CFR. 

To address these challenges, we propose GeoGS-CE, a light-of-sight (LoS)/non-LoS (NLoS)-aware 3D Gaussian framework for learning delay--beam channel priors. GeoGS-CE represents persistent NLoS scattering support with a scene-level Gaussian map, adapts Gaussian opacity, delay, and gain through a UE-location-conditioned deformer, and explicitly injects a virtual LoS path. A leakage-aware differentiable wireless renderer then projects the LoS and NLoS components into the delay--beam domain using signal-processing-based delay and beam kernels. The predicted delay--beam power spectrum serves as a structured covariance prior for online sparse-pilot linear minimum mean square error (LMMSE) channel estimation and short-term temporal prediction.

The main contributions are summarized as follows:
\begin{itemize}
    \item We propose GeoGS-CE, a geometry-conditioned 3D Gaussian framework that learns non-coherent delay--beam power priors for sparse-pilot high-mobility channel estimation.
    \item We design a LoS/NLoS-aware differentiable wireless renderer that combines Gaussian-modeled NLoS scatterers, an explicit virtual LoS path, and orthogonal frequency division multiplexing (OFDM) /array leakage-aware projection kernels towards the desired delay--beam power spectrum.
    \item We integrate the learned prior with online LMMSE estimation and lightweight temporal prediction to reconstruct and predict full-band, full-array CFRs from sparse pilots.
\end{itemize}

\section{System and Signal Model}

\begin{figure}
    \centering
    \includegraphics[width=0.8\linewidth]{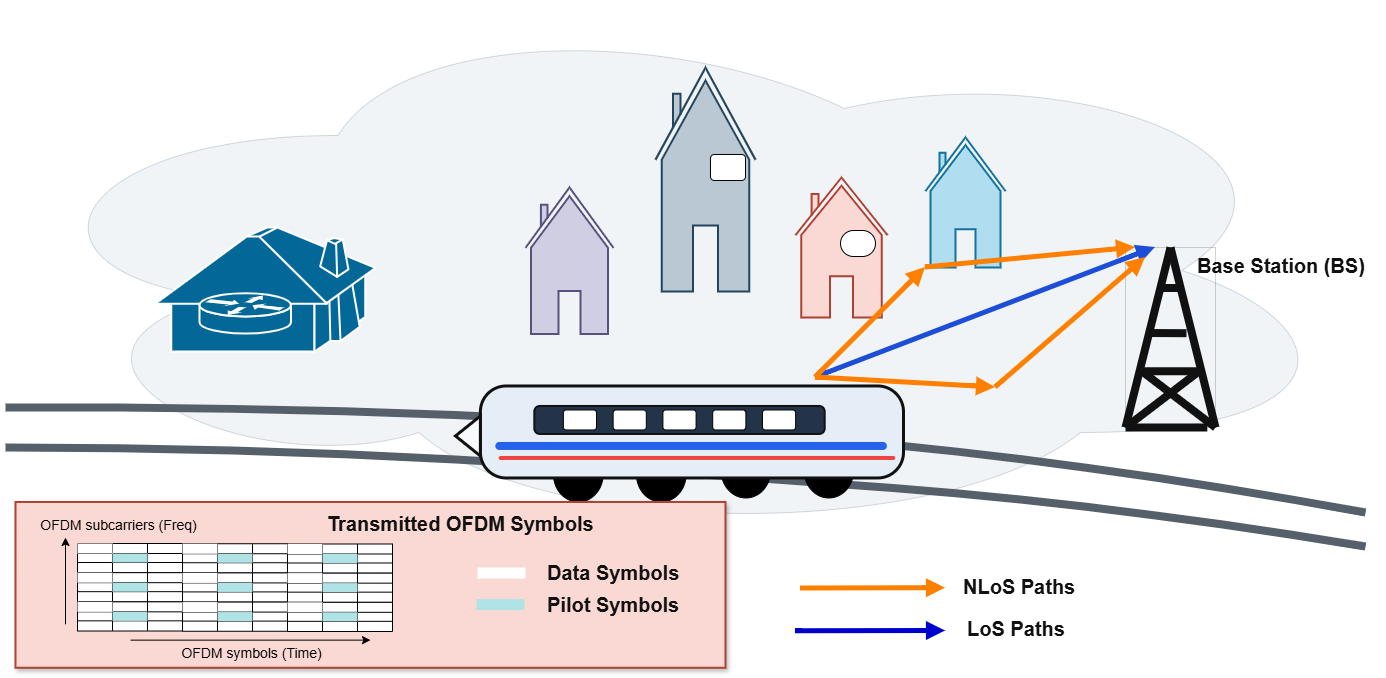}
    \caption{The high-mobility system diagram and the employed comb pilot pattern.}
    \label{fig:system_model}
\end{figure}
This section describes the high-mobility system model, the signal model and the target delay--beam power spectrum prior in Section II-A. We then formulate the learning objective for two stages in Section II-B.

\subsection{System and Radio Measurement Model}
Fig. \ref{fig:system_model} depicts the considered high-speed railway scenario. Assume a single-antenna user equipment (UE) moves along the track with a scheduled velocity, transmitting pilots to a base station (BS) equipped with an \(M\)-antenna uniform linear array (ULA).  The UE consistently transmits \(T\) OFDM symbols with \(N\) subcarriers to the BS. Following the 3GPP framework, we assume the insertion of comb pilots across OFDM subcarriers and symbols for channel reconstruction, as shown in the bottom-left of Fig. \ref{fig:system_model}. 

Specifically, let \(\mathcal{T}_{\mathrm{p}}\) denote the set of OFDM symbols with pilot measurements and \(\mathcal{N}_{\mathrm{p}}\) denote the pilot subcarrier set per measurement symbol. The received pilot observation at measurement symbol \(t \in \mathcal{T}_{\mathrm{p}}\) can be written as
\begin{equation}\label{eq:received_y}
    \mathbf{Y}_t = P_{\mathrm{tr}}\mathcal{S}_{\mathcal{P}}(\mathbf{H}_t) + \mathbf{W}_t \in \mathbb{C}^{N_{\mathrm{p}}\times M},
\end{equation}
where $P_{\mathrm{tr}}$ is the transmit power (assume identity transmit pilots without loss of generality). \(\mathbf{H}_t \in \mathbb{C}^{N\times M}\) is the full-band, full-array CFR at OFDM symbol \(t\), with \([\mathbf{H}_t]_{n,m}\) for \(n \in [0,\dots,N-1]\) and \(m \in [0,\dots,M-1]\) being the CFR on subcarrier \(n\) and BS antenna \(m\). The operator \(\mathcal{S}_{\mathcal{P}}(\cdot)\) selects the \(\mathcal{N}_{\mathrm{p}}\) rows of the CFR matrix corresponding to the pilot subcarriers with \({N}_{\mathrm{p}}=|\mathcal{N}_{\mathrm{p}}|\), and \(\mathbf{W}_t\) denotes additive white Gaussian noise. 

The objective is to estimate and predict \(\mathbf{H}_t\) for every OFDM symbol \(t\), which is intractable using a direct LMMSE approach since the observed pilots have much smaller dimensions than the unknown channel dimensions (i.e., \(N_{\mathrm{p}} \ll N\) and \(T_{\mathrm{p}} \ll T\)). However, by performing an inverse fast Fourier transform (IFFT) across the frequency domain and a fast Fourier transform (FFT) across the antenna domain, we can transform \(\mathbf{H}_t\) into its delay--beam representation as follows
\begin{equation}
    \mathbf{X}^{\mathrm{FULL}}_t
    =
    \mathbf{F}_N^{\mathrm{H}}
    \mathbf{H}_t
    \mathbf{F}_M
    \in \mathbb{C}^{N\times M},
\end{equation}
 where each nonzero element of \(\mathbf{X}^{\mathrm{FULL}}_t\) corresponds to an equivalent path component at delay bin \(\ell \in [0,\dots,N-1]\) and beam index \(u \in [0,\dots,M-1]\). 

In practice, valid paths only arrive at delays much smaller than the pilot count (\(L \ll N_p\)), and only a limited number of paths arrive from distinct angles. Therefore, the target delay--beam domain is a truncated version of \(\mathbf{X}^{\mathrm{FULL}}_t\) as follows
\begin{equation}\label{eq:X_truncate}
    \mathbf{X}_t
    =
    \mathbf{P}_{\mathcal{D}}\mathbf{X}^{\mathrm{FULL}}_t 
    = \mathbf{P}_{\mathcal{D}} \mathbf{F}_N^{\mathrm{H}}
    \mathbf{H}_t
    \mathbf{F}_M \in \mathbb{C}^{L\times M},
\end{equation}
where \(\mathbf{P}_{\mathcal{D}}\) selects the retained delay taps within the delay window \(\mathcal{D}\). Typically, we select \(\mathcal{D}=[-L, L]\) to accommodate the limited delay resolution of the FFT, which induces path delay leakage.

To facilitate subsequent channel estimation, we vectorize the received signal in Eq. \eqref{eq:received_y} with Eq. \eqref{eq:X_truncate}, which gives
\begin{align}\label{eq:yt}
    &\mathbf{y}_t
    = P_{\mathrm{tr}}\mathrm{vec}(\mathcal{S}_{\mathcal{P}}(\mathbf{H}_t)) =
    P_{\mathrm{tr}}\mathbf{A}_{\mathcal{P}}\mathbf{x}_t
    +
    \mathbf{w}_t,\\
    \mathrm{with}~&
    \mathbf{A}_{\mathcal{P}}
    =
    \mathbf{F}_M^{*}
    \otimes
    \left(
    \mathcal{S}_{\mathcal{P}}
    \mathbf{F}_N
    \mathbf{P}_{\mathcal{D}}^{\mathrm{T}}
    \right),
\end{align}
where \(\mathbf{x}_t = \mathrm{vec}(\mathbf{X}_t)\), \(\mathbf{w}_t = \mathrm{vec}(\mathbf{W}_t)\) and \(\mathbf{A}_{\mathcal{P}}\) is the sensing matrix induced by the inverse delay--beam transform and pilot subcarrier selection. 
 \(\otimes\) denotes the Kronecker product.

\subsection{Problem Formulation}
Upon this stage, we formulate the overall channel estimation problem of the two stages: (1) offline prior learning, and (2) online channel estimation incorporating predicted priors.

\subsubsection{Stage 1 -- Offline Prior Learning}
The objective of this stage is to learn a geometry-conditioned delay--beam power predictor \(f_{\Theta}\), which maps the UE position \(\mathbf{u}_t\) as follows
\begin{equation}
    \widehat{\mathbf{Q}}_t = f_{\Theta}(\mathbf{u}_t).
\end{equation}
whose parameters are decided to minimize the reconstruction loss \(\mathcal{L}\) over the dataset:
\begin{equation}
    \Theta^* = \arg\min_{\Theta} \mathbb{E} \left[ \mathcal{L} \left( \widehat{\mathbf{Q}}_t, \mathbf{Q}_t^{\mathrm{gt}} \right) \right].
\end{equation}
with $\mathbf{Q}_t^{\mathrm{gt}}$ being the ground-truth (GT) delay--beam power spectrum as follows
\begin{equation}
    \mathbf{Q}_t^{\mathrm{gt}}
    =
    |\mathbf{X}_t|^2
    \in \mathbb{R}_{+}^{L\times M},
\end{equation}
where \(|\cdot|^2\) is applied element-wise. The delay--beam power spectrum is relatively non-coherent, by characterizing the support and relative power distribution per propagation path in the delay--beam domain. Hence, it is more learning-friendly, getting rid of learning the instantaneous complex phases of the channel coefficients.

\textit{Remark:}  The above definition uses the clean full-band CFR and serves as the ideal delay--beam power target for evaluation. In practical offline training, when only noisy measurements are available, the supervision can be replaced by a surrogate target constructed from the observed CFR. Since GeoGS-CE learns the non-coherent power spectrum rather than instantaneous phases, the learned prior is less sensitive to measurement noise than coherent CFR supervision. 

\subsubsection{Stage 2 -- Online Channel Estimation and Prediction}
During the online operation, the predicted power spectrum \(\widehat{\mathbf{Q}}_t\) serves as a structural covariance prior for both channel estimation at measurement symbols and channel prediction at non-measurement symbols. Let \(\mathcal{Y} = \{\mathbf{y}_{\tau} \mid \tau \in \mathcal{T}_{\mathrm{meas}}\}\) denote the set of available sparse pilot observations. The online reconstruction can be formulated as a general mapping \(\mathcal{F}_{\mathrm{online}}\) that outputs the estimated delay--beam channel \(\widehat{\mathbf{x}}_t\) for any symbol \(t \in \{0, \dots, T_{\mathrm{slot}}-1\}\):
\begin{equation}
    \widehat{\mathbf{x}}_t = \mathcal{F}_{\mathrm{online}}\left( \mathcal{Y}, \widehat{\mathbf{Q}}_t \right).
\end{equation}
Specifically, \(\mathcal{F}_{\mathrm{online}}\) includes both LMMSE estimation at measurement symbols and temporal prediction at non-measurement symbols, as detailed in Section III-C.

\section{Proposed GeoGS-CE Framework}
\begin{figure*}
    \centering
    \includegraphics[width=0.8\linewidth]{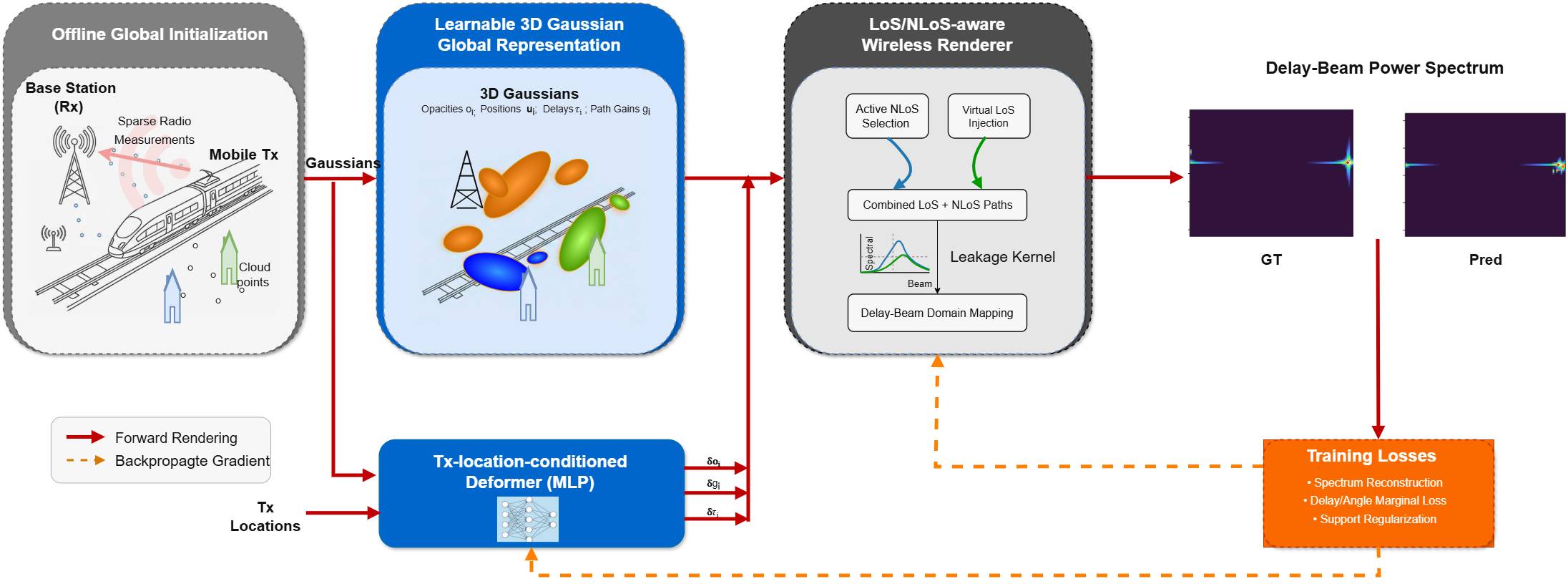}
    \caption{A scene-level set of 3D Gaussians models route-dependent NLoS scattering support, while a UE-location-conditioned deformer adapts their opacity, gain, and delay residuals to each receiver pose. The wireless renderer injects an explicit virtual LoS path and projects both LoS and NLoS components into the delay-beam domain. A learnable leakage layer models spectral spreading, and the predicted delay-beam spectrum is supervised by measurement-derived ground-truth spectra through reconstruction, marginal, and support-aware losses.}
    \label{fig:pipeline}
\end{figure*}
In this section, we describe the proposed two-stage channel estimation framework with sparse measurements in high-mobility scenarios. We first provide a framework overview in Section~III-A, then present the offline delay--beam prior learning stage in Section~III-B and the online prior-aided estimation stage in Section~III-C.

\subsection{Overview}
The proposed framework follows an offline--online pipeline for prior-aided channel estimation in high-mobility railway scenarios. The key idea is to decouple the relatively stable geometry-induced scattering support, evaluated by the delay--beam power spectrum, from rapidly varying coherent channel coefficients. Upon achieving the strong delay--beam power spectrum prior, we obtain the random phase information through low-complexity LMMSE incorporating with sparse radio measurements.

\textbf{In the offline stage}, we learn a scene-level delay--beam prior from training measurements. As illustrated in Fig.~\ref{fig:pipeline}, the offline model consists of three main components:
\begin{itemize}
    \item A scene-level 3D Gaussian channel map, which represents persistent NLoS scattering support along the railway segment.
    \item A channel-UE-location-conditioned deformer, which predicts bounded residual corrections to Gaussian opacity, gain, and delay according to the query link geometry and motion state.
    \item A LoS/NLoS-aware differentiable wireless renderer, which injects an explicit virtual LoS component and projects both LoS and Gaussian-induced NLoS components into the delay--beam domain using fixed OFDM delay and array beam kernels. These kernels account for practical delay/beam leakage while keeping the projection physically constrained and interpretable.

\end{itemize}

\textbf{In the online stage}, the rendered delay--beam power spectrum serves as a structured prior for sparse-pilot CFR estimation and short-term channel prediction.

\subsection{Offline delay--beam Prior Training} 

To accelerate convergence and improve physical consistency, the 3D Gaussians are initialized by using a measurement-guided  strategy, where dominant delay--beam components are extracted from surrogate CIR measurements from available training CFR measurements and back-projected into candidate scattering locations generating from the geometry cloud points. These components initialize Gaussian positions, opacities, delay residuals, and radio gains, providing a warm start for the differentiable rendering pipeline.

\subsubsection{Scene-Level 3D Gaussian Representation}
In the proposed framework, each 3D Gaussian primitive is regarded as a potential environmental scatterer, that facilitates mimicking a potential NLoS propagation path in the railway scenario. Similar to conventional 3D Gaussian representations, for the $i$-th ($i\in [1,...,G]$) primitive with $G$ being the number of Gaussian primitives, it has a central position $\boldsymbol{\mu}_i\in\mathbb{R}^3$, a volumetric support $\boldsymbol{\Sigma}_i$, and an opacity-like activation weight $o_i$ that controls the visibility (existence) of the represented NLoS path.  These attributes describe where a potential scatterer is located, how spatially the scattering support contributes to a given link. 

However, wideband wireless propagation requires additional radio attributes beyond geometry, delay-beam power spectrum is not only characterized by scatterers' spatial support, but also by their propagation delay and path gain. Therefore, we augment each Gaussian primitive with a learnable delay residual and a learnable radio gain. The scene-level Gaussian channel map is defined as
\begin{equation}
    \mathcal{G}_{\Theta}
    =
    \left\{
    \boldsymbol{\mu}_i,
    \boldsymbol{\Sigma}_i,
    o_i,
    \Delta\tau_i,
    g_i
    \right\}_{i=1}^{N_G},
\end{equation}
where  $\Delta\tau_i$ is a learnable delay residual, and $g_i$ is a learnable base path gain.

For a given receiver sample, the Gaussian center determines the geometric BS-side direction and the coarse propagation distance, while $\Delta\tau_i$ compensates for delay mismatch caused by simplified geometry and unresolved propagation effects. The gain $g_i$ controls the radio strength of the corresponding scattering component.

\subsubsection{UE-location-conditioned Gaussian Deformer}
Although the Gaussian map is globally shared, the visibility and effective contribution of each scatterer vary with the receiver position and motion state. We therefore introduce a UE-location-conditioned deformer that predicts bounded residual corrections for each Gaussian. For the $k$-th query sample, the deformer takes the mobile receiver position $\mathbf{u}_k$ and the $i$-th Gaussian primitives as input:
\begin{equation}
    \mathbf{z}_{i,k}
    =
    \left[
    \boldsymbol{\mu}_i,
    \mathbf{u}_k
    \right],
\end{equation}
 The deformer outputs
\begin{equation}
    (\delta o_{i,k},\delta\tau_{i,k},\delta g_{i,k})
    =
    d_{\Phi}(\mathbf{z}_{i,k}).
\end{equation}
Finally, the effective UE-location-conditioned attributes of the scene-level Gaussian representatives are
\begin{align}
    \tilde{o}_{i,k}
    &=
    \sigma\left(o_i+\eta_o\tanh(\delta o_{i,k})\right),\\
    \tilde{\tau}_{i,k}
    &=
    \Delta\tau_i+\eta_{\tau}\tanh(\delta\tau_{i,k}),\\
    \tilde{g}_{i,k}
    &=
    \mathrm{softplus}\left(g_i+\eta_g\tanh(\delta g_{i,k})\right),
\end{align}
where $\eta_o$, $\eta_{\tau}$, and $\eta_g$ bound the magnitude of the residual corrections. The deformer is initialized to produce near-zero residuals, so the model starts from the global Gaussian map and gradually learns sample-specific modulation.

\subsubsection{LoS/NLoS-Aware Differentiable Wireless Renderer}
For each UE channel sample, the deformed Gaussians are interpreted as candidate NLoS paths. The delay of the $i$-th Gaussian-induced NLoS component is modeled as
\begin{equation}
    \tau_{i,k}^{\mathrm{NLoS}}
    =
    \|\boldsymbol{\mu}_i-\mathbf{b}\|_2
    /{c}
    +
    \tilde{\tau}_{i,k},
\end{equation}
where $c$ is the speed of light. The residual delay term absorbs the remaining geometry- and propagation-dependent delay not captured by the BS-to-Gaussian distance alone.

To prevent the dominant direct path from being redundantly explained by NLoS Gaussians, we explicitly inject a virtual LoS component:
\begin{equation}
    \tau_{k}^{\mathrm{LoS}}
    =
    {\|\mathbf{u}_k-\mathbf{b}\|_2}/{c}.
\end{equation}
The LoS and NLoS components are combined into one path set and projected by the same delay--beam operator.

The rendered delay--beam power spectrum is given by
\begin{equation}
    \widehat{Q}_{k}(\ell,b)
    =
    \sum_{p\in\mathcal{P}_k}
    w_{p,k}
    K_{\tau}(\tau_{\ell}-\tau_{p,k})
    K_{b}(\nu_b-\nu_{p,k}),
\end{equation}
where \(\mathcal{P}_k\) contains the explicit LoS path and the selected Gaussian-induced NLoS paths, \(w_{p,k}\), \(\tau_{p,k}\), and \(\nu_{p,k}\) denote the path power, delay, and BS-side beam coordinate, respectively.

The delay kernel is determined by the finite-subcarrier IFFT response. For \(N\) subcarriers with spacing \(\Delta f\), an off-grid path delay \(\tau\) contributes to delay tap \(\ell\) according to
\begin{equation}
    K_{\tau}(\tau_{\ell}-\tau)
    =
    \left|
    \frac{1}{N}
    \sum_{n=0}^{N-1}
    e^{-j2\pi n\Delta f(\tau_{\ell}-\tau)}
    \right|^2.
\end{equation}
Similarly, the beam kernel is induced by the finite ULA aperture:
\begin{equation}
    K_b(\nu_b-\nu)
    =
    \left|
    \frac{
    \mathbf{a}(\nu_b)^{H}\mathbf{a}(\nu)
    }{
    \|\mathbf{a}(\nu_b)\|_2\|\mathbf{a}(\nu)\|_2
    }
    \right|^2.
\end{equation}
These fixed kernels model the delay and beam spreading caused by finite OFDM bandwidth and finite array aperture. Therefore, the renderer remains differentiable with respect to Gaussian path attributes, while the projection itself is constrained by signal-processing physics rather than learned freely from data.

\subsubsection{Offline Training Objective}
Given the GT delay-beam power spectrum \(\mathbf{Q}_k^{\mathrm{gt}}\), the primary reconstruction loss matches the rendered and target spectra as follows
\begin{equation}
    \mathcal{L}_{\mathrm{spec}}
=
\frac{1}{|\Omega|}
\sum_{\ell,b}
\left(
\bar Q_k[\ell,b]-\bar Q_k^{\mathrm{gt}}[\ell,b]
\right)^2,
\end{equation}
where $\Omega$ is the weight.

However, element-wise spectral matching alone can be sensitive to small delay or beam misalignments. We therefore introduce marginal losses along the delay and beam axes as a smooth warm-up:
\begin{align}
\mathcal{L}_{\mathrm{marg}}
=&
\lambda_d \sum_\ell
\left(\bar q_k^{d}[\ell]-\bar q_k^{d,\mathrm{gt}}[\ell]\right)^2
+\\
&\lambda_b \sum_b
\left(\bar q_k^{b}[b]-\bar q_k^{b,\mathrm{gt}}[b]\right)^2,
\end{align}
where $\bar{\mathbf q}_k^d$ and $\bar{\mathbf q}_k^b$ are the unit-sum normalized delay and beam marginals of $\widehat{\mathbf Q}_k$, and $\bar{\mathbf q}_k^{d,\mathrm{gt}}$ and $\bar{\mathbf q}_k^{b,\mathrm{gt}}$ are defined analogously from $\mathbf Q_k^{\mathrm{gt}}$. $\lambda_d$ and $\lambda_b$ are the respective weights. These terms stabilize training by preserving global delay and beam statistics even when local spectral bins are slightly shifted.

Finally, we use support-aware regularization to improve the precision and recall of the learned path support:
\begin{equation}
    \mathcal{L}_{\mathrm{sup}}
    =
    \lambda_{\mathrm{false}}\mathcal{L}_{\mathrm{false}}
    +
    \lambda_{\mathrm{recall}}\mathcal{L}_{\mathrm{recall}}
    +
    \lambda_{\mathrm{LoS}}\mathcal{L}_{\mathrm{LoS-false}}.
\end{equation}
Herein, \(\mathcal{L}_{\mathrm{false}}\) suppresses predicted energy outside the target support, \(\mathcal{L}_{\mathrm{recall}}\) encourages the renderer to cover weak but valid NLoS support, and \(\mathcal{L}_{\mathrm{LoS-false}}\) penalizes Gaussian-induced NLoS energy around the LoS region, to avoid repetitive conflict with the  virtual LoS component.

The total offline objective is then
\begin{equation}
    \mathcal{L}
    =
    \lambda_{\mathrm{spec}}\mathcal{L}_{\mathrm{spec}}
    +
    \lambda_{\mathrm{marg}}\mathcal{L}_{\mathrm{marg}}
    +
    \mathcal{L}_{\mathrm{sup}}.
\end{equation}

\subsection{Online Prior-Aided LMMSE and Temporal Prediction}

During online operation, the trained GeoGS-CE prior is queried with the current link geometry to predict a delay--beam power spectrum \(\widehat{\mathbf{Q}}_t\). We convert this non-coherent spectrum into a diagonal covariance prior for the delay--beam channel vector:
\begin{equation}
    \mathbf{C}_{x,t}
    =
    \rho_t
    \mathrm{diag}
    \left(
    \mathrm{vec}
    \left(
    \widehat{\mathbf{Q}}_t+\epsilon_q
    \right)
    \right),
\end{equation}
where \(\epsilon_q\) is a small power floor and \(\rho_t\) is a scalar power calibration factor estimated from pilot energy or large-scale fading information.

Given the sparse pilot observation \(\mathbf{y}_t\), the prior-aided LMMSE estimate is
\begin{equation}
    \widehat{\mathbf{x}}_t
    =
    \mathbf{C}_{x,t}
    \mathbf{A}_{\mathcal{P}}^{\mathrm{H}}
    \left(
    \mathbf{A}_{\mathcal{P}}
    \mathbf{C}_{x,t}
    \mathbf{A}_{\mathcal{P}}^{\mathrm{H}}
    +
    \sigma_w^2\mathbf{I}
    \right)^{-1}
    \mathbf{y}_t .
\end{equation}
The estimated delay--beam vector is then transformed back to obtain the full-band, full-array CFR.

For OFDM symbols without pilot measurements, we adopt a lightweight causal temporal predictor. Let \(t_{\mathrm{ref}}\) denote the most recent measurement symbol. The delay--beam state is predicted as
\begin{equation}
    \widehat{\mathbf{x}}_{t}
    =
    \widehat{\alpha}^{t-t_{\mathrm{ref}}}
    \widehat{\mathbf{x}}_{t_{\mathrm{ref}}},
\end{equation}
where \(\widehat{\alpha}\in\mathbb{C}\) is estimated from recent measurement-aided posterior states. Once a new measurement symbol arrives, the predicted state is refreshed by the LMMSE update.

\section{Simulations}

This section evaluates GeoGS-CE on a generated high-speed railway channel dataset. We first describe the simulation setup, then evaluate the offline delay--beam prior prediction and the online sparse-pilot CFR reconstruction performance.

\subsection{Simulation Setup}

We consider a \(500\,\mathrm{m}\) segment of the Guangshen high-speed railway and generate channel measurements using Sionna RT. A single-antenna UE moves along the railway track at a constant velocity of \(350\,\mathrm{km/h}\), while the BS is equipped with a \(100\times 1\) ULA. The OFDM system uses \(N=1024\) subcarriers and \(T_{\mathrm{slot}}=14\) OFDM symbols per slot. Sparse pilot measurements are inserted on 4 OFDM symbols per slot, with \(N_p=72\) comb pilot subcarriers per measurement symbol. Unless otherwise specified, the SNR is set to \(20\,\mathrm{dB}\).

We generate approximately \(15{,}000\) sequential CFR measurements along the railway route. A small subset (around $300$) of channels is used for offline prior learning, while the remaining positions are reserved for testing and prediction. The reconstruction quality is evaluated by the full-band, full-array CFR NMSE defined by $\mathrm{NMSE}_t
    =
    10\log_{10}
    \frac{
    \|\widehat{\mathbf{H}}_t-\mathbf{H}_t\|_F^2
    }{
    \|\mathbf{H}_t\|_F^2
    }.$

We compare GeoGS-CE with the following baselines:
\begin{itemize}
    \item \textbf{Zero-prior LMMSE}: a low-complexity baseline that performs LMMSE estimation with an uninformative covariance prior and uses channel holding for non-measurement symbols.
    \item \textbf{WRF-GS$^+$ prior}: a strong  3D-GS-based learning baseline. Its predicted prior is combined with the same LMMSE estimator and scalar temporal predictor for fair comparison.
    \item \textbf{LoS-seeded OMP prior}: a compressed-sensing baseline initialized with strong LoS information, followed by the same temporal prediction rule.
    \item \textbf{Genie prior}: an oracle lower bound that uses the ground-truth delay--beam power spectrum as the covariance prior.
\end{itemize}

\subsection{Offline Evaluation of Delay--Beam Prior Prediction}

\begin{figure}
    \centering
    \includegraphics[width=0.8\linewidth]{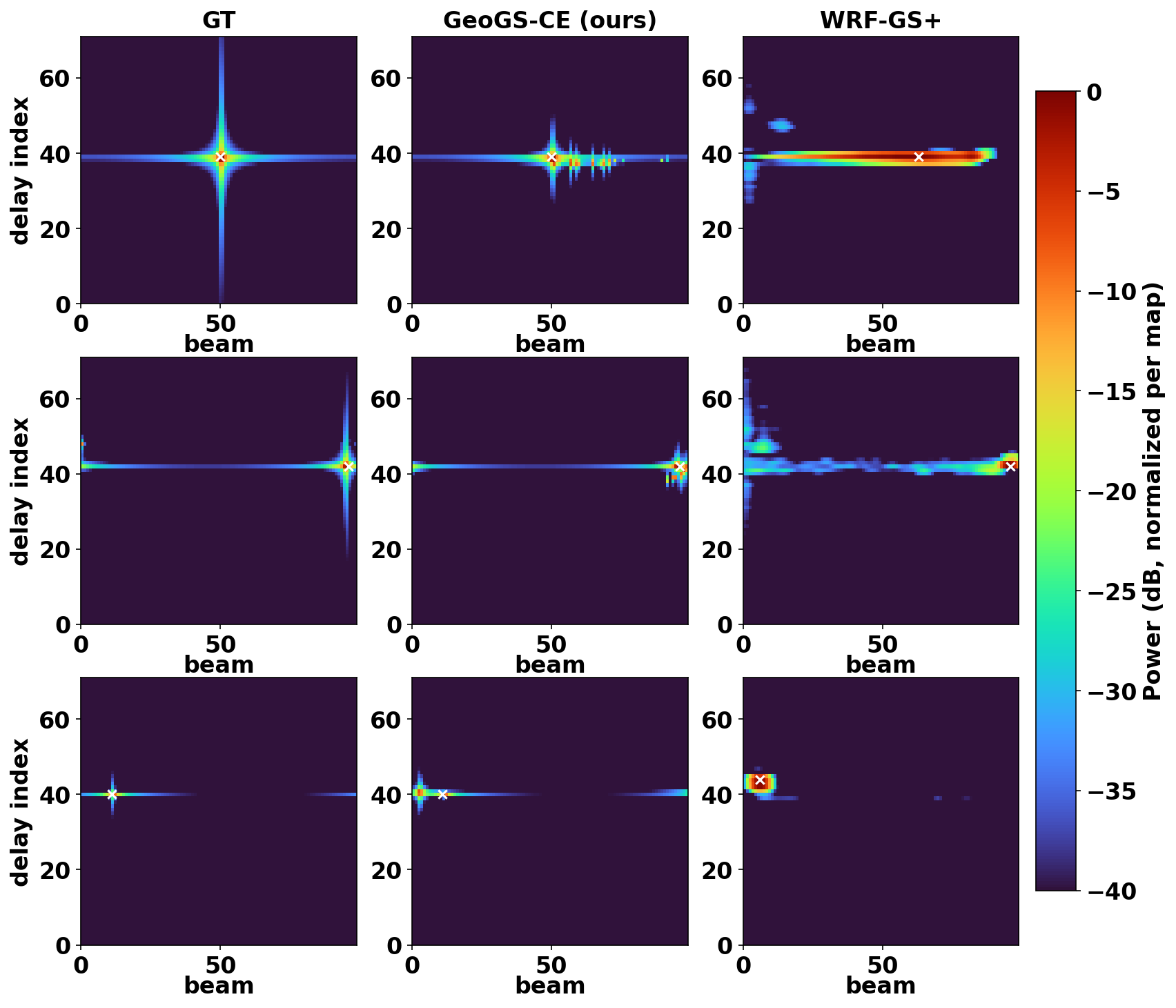}
    \caption{Offline delay--beam power spectrum prediction at held-out railway positions. Left: measurement-derived target. Middle: proposed GeoGS-CE prior. Right: baseline prediction. GeoGS-CE better localizes the dominant support and suppresses false delay--beam leakage.}
    \label{fig:pdp_spectrum}
\end{figure}

We first evaluate the performance of the offline model in predicting the non-coherent delay--beam power spectrum at unseen railway positions. Fig.~\ref{fig:pdp_spectrum} compares the GT, the proposed GeoGS-CE prior prediction, and the WRF-GS$^+$ prediction, and shows that GeoGS-CE better preserves the localized LoS/NLoS support and suppresses broad false-support leakage. In contrast, the WRF-GS$^+$ baseline tends to produce excessive horizontal spreading and more severely misplaced scattering components, indicating that a purely learned scene representation without the LoS/NLoS-structure-aware rendering structure is less effective at capturing sparse delay -beam support.

\subsection{Online Sparse-Pilot Channel Estimation and Prediction}
\begin{figure}[t]
    \centering
    \begin{subfigure}[t]{0.48\linewidth}
        \centering
        \includegraphics[width=\linewidth]{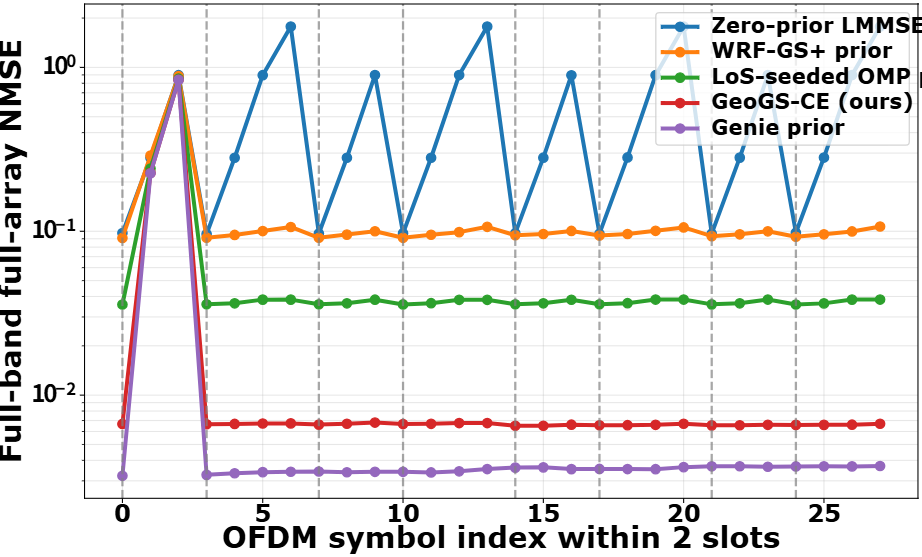}
        \caption{Median NMSE}
        \label{fig:nmse_median}
    \end{subfigure}
    \hfill
    \begin{subfigure}[t]{0.48\linewidth}
        \centering
        \includegraphics[width=\linewidth]{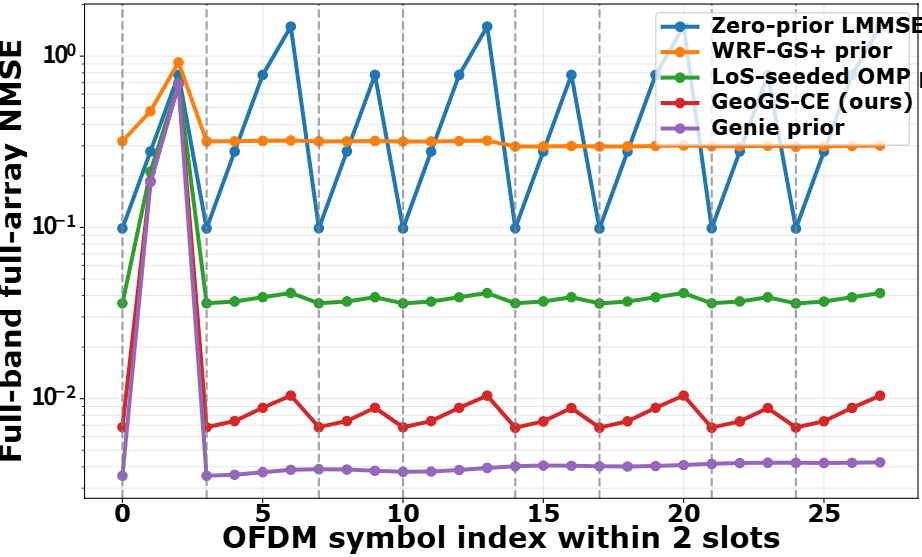}
        \caption{Mean NMSE}
        \label{fig:nmse_mean}
    \end{subfigure}
    \caption{Full-band, full-array CFR NMSE over two consecutive slots, with (a) median NMSE, and (b) mean NMSE. The first 4 OFDM symbols perform channel holding for all schemes to initialize the temporal predictor calculation. }
    \label{fig:nmse_online}
\end{figure}
We further evaluate full-band, full-array CFR reconstruction and prediction from sparse pilot observations. Fig.~\ref{fig:nmse_online} reports both the median (a) and mean (b) NMSE over test snapshots across two consecutive slots.  The vertical dashed lines mark OFDM symbols with pilot measurements, where the channel state is refreshed by the prior-aided LMMSE estimator. Other remaining symbols without measurements are predicted by the scalar temporal predictor as detailed in Section III-C.

As shown in Fig.~\ref{fig:nmse_median}, GeoGS-CE achieves the best nmse performance among all practical methods and closely follows the genie-prior bound. This indicates that the learned delay--beam power prior captures most of the dominant non-coherent channel structure needed for sparse-pilot reconstruction. Compared with zero-prior LMMSE, the proposed prior significantly reduces the reconstruction error, especially at non-measurement symbols where channel holding quickly degrades.

The mean NMSE in Fig.~\ref{fig:nmse_mean} further demonstrates the robustness of GeoGS-CE. After the early transition symbols, GeoGS-CE rapidly returns to a low-error regime after measurement updates and maintains a clear performance advantage over WRF-GS$^+$ and LoS-seeded OMP. The gap between the median and mean curves between those of Genie lower bounds remains small for GeoGS-CE compared with WRF-GS$^+$, suggesting that the learned geometry-aware prior is less sensitive to difficult channel snapshots and false-support predictions.

\begin{table}[t]
\centering
\caption{Ablation study on channel-estimation NMSE (mean and medium regimes). }
\label{tab:ablation_target4}
\begin{tabular}{lcc}
\hline
Method & Mean NMSE (dB)$\downarrow$  & Medium NMSE (dB)$\downarrow$  \\
\hline
\textbf{Proposed} & \textbf{-14.07} & \textbf{-20.84}\\
No virtual LoS & -13.46 & -18.32 \\
No init + No virtual LoS & -12.32 & -15.47 \\
No leakage kernel & 9.78 & 9.76 \\
\hline
\end{tabular}
\end{table}

Table~\ref{tab:ablation_target4} then evaluates the contribution of key design components to channel-estimation accuracy. The baseline achieves the best NMSE in both regimes, while removing virtual LoS causes a clear degradation, and jointly removing measurement data initialization and virtual LoS further amplifies the loss. The no-leakage-kernel setting exhibits a severe collapse NMSE, indicating that the leakage/rendering kernel is essential for stable and physically consistent reconstruction.

\section{Conclusion}

This paper proposed GeoGS-CE, a two-stage channel estimation and prediction framework for sparse-pilot high-mobility wideband massive-MIMO systems. The offline stage learns a geometry-conditioned, non-coherent delay--beam power prior using a LoS/NLoS-aware 3D Gaussian representation and a differentiable wireless renderer. The online stage converts the predicted delay--beam spectrum into a structured covariance prior for LMMSE channel estimation and lightweight temporal prediction. Simulations on a Sionna-RT-generated Guangshen high-speed railway scenario show that GeoGS-CE substantially improves full-band, full-array CFR reconstruction over pilot-only, compressed-sensing, and baseline 3D-GS prior methods, while approaching the genie-prior bound.
\bibliographystyle{IEEEtran}

\bibliography{IEEEabrv, reference}
\end{document}